\documentclass[a4paper]{jpconf}

\usepackage{graphicx}
\usepackage{aasjournals}
\usepackage{hyperref}
\graphicspath{{./}{Figures/}}

\begin{document}
\title{Challenges Modeling the Low-Luminosity Type Iax Supernovae}

\author{C Feldman$^{1,2}$, E Eisenberg$^3$, D M Townsley$^4$ and A C Calder$^{1,2}$}

\address{$^1$ Department of Physics and Astronomy, 
Stony Brook University, Stony Brook, NY 11794-3800, USA}
\address{$^2$ Institute for Advanced Computational Science,
Stony Brook University, Stony Brook, NY 11794-5250, USA}
\address{$^3$ Department of Physics, Yale University, New Haven, CT 06520, USA}
\address{$^4$ Department of Physics and Astronomy, The University of Alabama, Tuscaloosa, AL 35487-0324, USA}

\ead{catherine.feldman@stonybrook.edu}

\begin{abstract}
Numerical models allow the investigation of phenomena that cannot exist in a laboratory. Computational simulations are therefore essential for advancing our knowledge of astrophysics, however, the very nature of simulation requires making assumptions that can substantially affect their outcome. Here, we present the challenges faced when simulating dim thermonuclear explosions, Type Iax supernovae. This class of dim events produce a slow moving, sparse ejecta that presents challenges for simulation. We investigate the limitations of the equation of state and its applicability to the expanding, cooling ejecta. We also discuss how the``fluff'', i.e.\ the low‐density gas on the grid in lieu of vacuum, inhibits the ejecta as it expands. We explore how the final state of the simulation changes as we vary the character of the burning, which influences the outcome of the explosion. These challenges are applicable to a wide range of astrophysical simulations, and are important to discuss and overcome as a community.
\end{abstract}

\section{Introduction}

Numerical modeling is the bridge between theory and observation, allowing experimentation in a controlled virtual laboratory. Astrophysics heavily relies on numerical modeling, as it depends on processes than span length scales from the quantum level to galaxies, and timescales of milliseconds to billions of years and most environments are difficult to reproduce in the laboratory. With simulations, we can collide galaxies, explode stars, and form planets, which allows us to test our theories of the Universe. But as British statistician George Box reminds us, ``All models are wrong, but some are useful.'' Choices for initial conditions, simplifying assumptions, and approximations of large- or small-scale processes need to be made in any numerical model for it to be computationally feasible. These choices can determine the entire outcome of a simulation, and it is therefore important to know a simulation's limitations and quirks, and determine quantitatively how the outcome depends on them. Despite their importance, the reasoning and selection of these choices are more often than not passed over in the literature, but are nonetheless considered common knowledge in the field.

Our application, thermonuclear supernovae, is no different. These events, known as Type Ia supernovae (SNe Ia), are created when a white dwarf (WD) star gains enough mass to reignite nuclear fusion and subsequently explode. While advances have been made in both observational and computational capabilities, both the characteristics of the initial system and the method of explosion are still a matter of debate \cite{Liu_2023}. Modeling SNe Ia involves processes from sub-centimeter scales (the initial ``flame'', where nuclear burning runs away) to thousands of kilometers (the size of the star). No simulation can simultaneously resolve both of these scales, so approximations must be made, usually by the introduction of a burning model that is tuned to reproduce the energy release of small-scale flame experiments. The initial star is often parameterized, as is the location and size of the ignition.  

This study focuses on the dimmest SNe Ia, a subclass known as Type Iax supernovae (SNe Iax). These events are thought to occur when a WD undergoes a deflagration (subsonic burning) without a subsequent detonation (supersonic burning) \cite{Foley_2013}. The lack of a vigorous explosion makes these simulations especially sensitive to initial conditions. This paper seeks to explicitly lay out and discuss the challenges we faced while performing an exploratory study of SNe Iax. Section \ref{sec:methods} briefly describes our simulation; Section \ref{sec:challenges} describes the challenges we faced; and Section \ref{sec:summary} summarizes our findings and considers the next steps.

\section{Methods} \label{sec:methods}
Our simulation for Type Iax events is briefly described here. Full details about our exploratory 2D deflagration study and its results can be found in \cite{Feldman_2023}.

To simulate the deflagration, we use an adapted version of FLASH, a highly parallelizable, multi-scale, multi-physics simulation instrument used to model the dynamics and evolution of astrophysical and high-energy phenomena \cite{Fryxell_2000}. FLASH uses the PARAMESH adaptive mesh library to decompose a volume into grid blocks where features of interest can have a higher resolution (i.e.\ have smaller sized blocks) than the rest of the grid. This ability allows FLASH to accurately model radiation magneto-hydrodynamic flows while putting heavy computation only where needed. FLASH's modular structure allows it to be extended to model different problems. For our study, we use FLASH version 4.6.2 with a custom burning model developed specifically for thermonuclear supernovae \cite{Townsley_2016}.

Our investigative 2D simulations of a deflagration are performed in FLASH on a cylindrically symmetric grid. First, a realistic progenitor \cite{Brooks_2017} created with MESA \cite{Paxton_2011} was mapped to the grid in FLASH \cite{Willcox_2016}. FLASH's hydrodynamics solver requires gas on the entire grid so, instead of the vacuum of space, low density ``fluff'' material was placed everywhere outside of the star. The initial conditions of the fluff were chosen with care, as the fluff can introduce artifacts and affect the results of the simulation. A deflagration was then artificially ignited by placing a circular 150 km fully-burned region at the temperature peak at the center of the star. To create conditions for plumes to form, random sinusoidal perturbations were placed on the circumference of this burned region. Thirty sets of random perturbations, each of which is called a ``realization'', were used to create a sample. 

The equation of state (EOS) used for our supernova application is the Helmholtz EOS from \cite{TimmesAndSwesty_2000} with Coulomb corrections, that is distributed with FLASH \cite{Fryxell_2000}. This EOS is meant to describe a degenerate, fully ionized, electron-positron plasma such as that inside a WD. The hydrodynamics solver uses the split piecewise parabolic method; gravity uses a multipole Poisson solver; the deflagration front is artificially thickened to a 4-grid-zone wide layer and tracked by a scalar variable; and nuclear burning is approximated by 3 distinct stages tracked by scalar progress variables -- the first of which (C-burning) is evolved with a modified advection-diffusion-reaction (ADR) scheme and the second (C-burning ash to Si-group elements) and third (Si-group to Fe-group elements) of which are evolved at tuned timescales -- plus a dynamic final burned state in nuclear statistical equilibrium. 

Each of the thirty realizations were run until 20 s. As found in other studies, the deflagration alone is not strong enough to unbind the entire WD \cite{Kromer_2015, Lach_2021}. The resulting system is a hot, dense, gravitationally bound remnant surrounded by sparse plumes of ash. Some parts of the plumes have enough kinetic energy to escape the pull of the remnant; this material will be ejected from the system, while the remaining material will fall back onto the remnant. Our simulation produces $^{56}$Ni yields and ejecta velocities that are in agreement with observations, but ejecta and kinetic energy yields that are 10 times smaller than observations suggest. These findings propel us to continue work in a 3D simulation that can model the interaction between the flame and background turbulence \cite{Jackson_2014}.

\section{Simulation Challenges} \label{sec:challenges}
This paper focuses on the challenges faced by simulating this slow evolving phenomenon that produces low-density, low-temperature material in a partially gravitationally bound environment. These challenges include reaching the limitations of the EOS (Section \ref{sec:eos}), slowdown of ejecta caused by the fluff (Section \ref{sec:fluff}), results dependent on our choice of included or excluded physics modules (Section \ref{sec:reignition}), and variability of the results due to variation of the initial conditions, in this case the initial flame radius (Section \ref{sec:initflame}). While discussed here in the context of our Type Iax application, these challenges are applicable to many simulations involving a grid-based approach to stellar phenomena.

\subsection{Equation of State} \label{sec:eos}

\begin{figure}[ht]
\begin{center}
\includegraphics[scale=0.8]{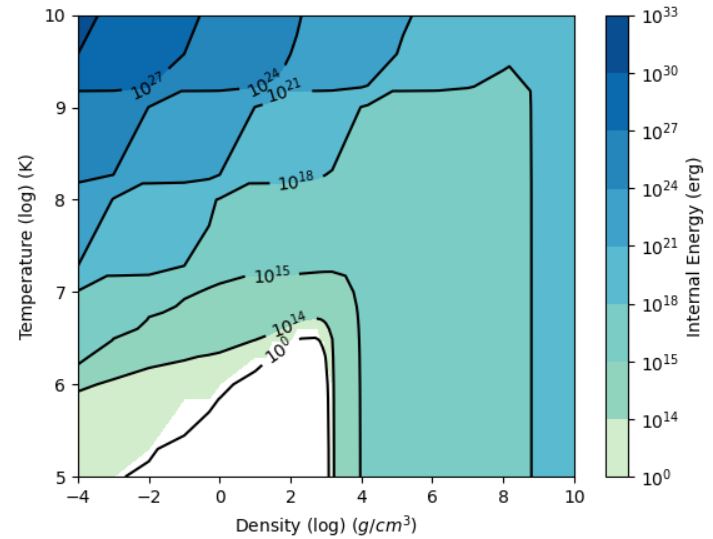}
\end{center}
\caption{\label{fig:eos_contour} Contour plot of internal energy in $\rho$-T space. Data was collected by sampling from the FLASH EOS using density and temperature as inputs. The white triangular region shows where the internal energy is negative, and therefore the EOS is invalid.}
\end{figure}

Our simulation would sometimes return a negative pressure and internal energy at the edges of the rising and expanding plumes. Finding that the Coulomb corrections would overwhelm the pressure in these cases, we decided to investigate the EOS. We called the FLASH EOS with intervals in density-temperature ($\rho$-T) space as input, and examined the resulting pressure and internal energy. The latter is plotted in Figure \ref{fig:eos_contour} as a function of the density and temperature.

At temperatures lower than 10$^7$ K, as the density decreases from 10$^4$ to 10$^3$ g/cm$^3$ the internal energy sharply drops by over ten orders of magnitude. The internal energy then becomes negative in a triangular-shaped region in $\rho$-T space that begins at densities on the order of 10$^3$ g/cm$^3$ and temperatures on the order of 10$^6$ K. At even lower densities, the internal energy remains negative with a decrease in temperatures. This triangular region of negative pressure and internal energy can also be seen in the newest version of the Helmoltz EOS, denoted the Skye EOS \cite{Skye_EOS}. In this region, the assumption that the gas is fully ionized breaks down, and this EOS becomes invalid.

\begin{figure}[h]
\begin{center}
\includegraphics[scale=0.2]{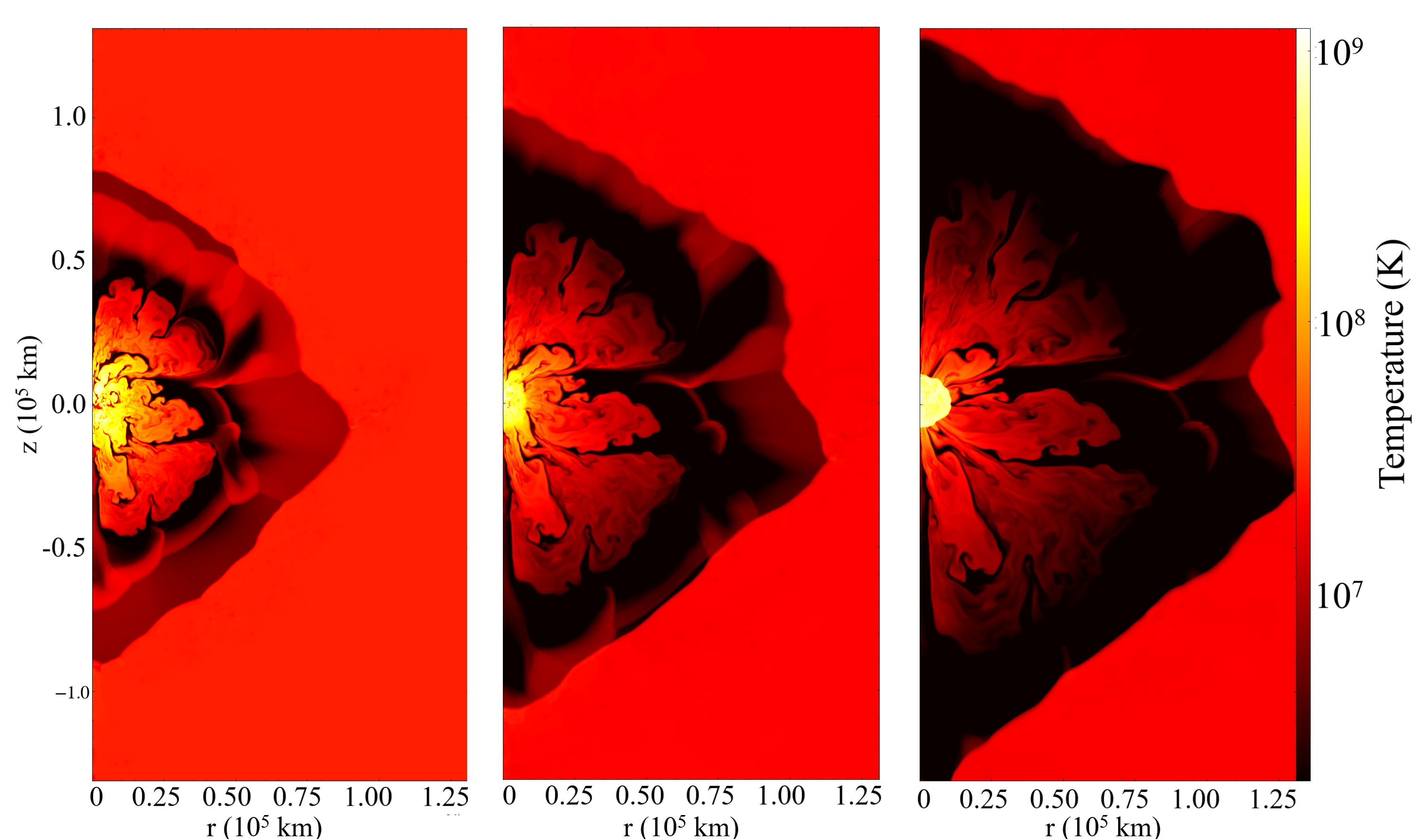}
\end{center}
\caption{\label{fig:temp_floor} Temperature of our simulation at (from left to right) 10.1 s, 15 s, and 19.8 s. Regions of low temperature can be seen in front of the expanding plumes of burned material. The temperature floor of 2 $\times$ 10$^6$ K is shown in black on the plots, and is reached as the ejecta expand and cool.}
\end{figure}

Our ``workaround'' to this issue was to set the temperature floor to the relatively high value of 2 $\times$ 10$^6$ K to prevent the material from entering the invalid region of the EOS. As can be seen in Figure \ref{fig:temp_floor}, the edges of the ejecta and the surrounding material all reach and are held at the temperature floor. As the simulation continues, more and more material is held at the temperature floor, which means that energy was added to the simulation in these regions. This artificial increase of temperature, and therefore pressure and internal energy, did not significantly affect our results. In the future, an internal energy floor could be used instead, as long as the results also remain unaffected by the artificial increase in energy.

The next step is to determine why our simulation ventures into this problematic area of $\rho$-T space. Does the expanding ejecta naturally fall into this region, implying a physically motivated cause that will require the coupling of the current EOS to another? Or, do the numerics of the FLASH solvers, perhaps when coupled to our burning model, artificially lead the solution astray?

\subsection{Fluff-ejecta Interaction} \label{sec:fluff}

The fluff is another feature of the simulation that inhibits running out to long (\textgreater 20 s) times. As the star expands, it collides and interacts with this low density gas. As the ejecta plows through the fluff, a reverse shock is sent into the ejecta, slowing it down. This effect has been seen in DDT and double detonation simulations \cite{Townsley_2019} and is also shown here, where the energy release and expansion velocities are much lower.

The properties (density, temperature, pressure, etc.) of the fluff must therefore be set to prevent an imbalance at the fluff-star boundary. To prevent a composition gradient, the composition of the fluff is set to match the composition of the outermost zone of the star. Here, the pressure in the fluff is set 25\% higher than the pressure at the edge of the star, to prevent any shocks at the beginning of the simulation where the outer edge of the star expands in reaction to the sudden introduction of the flame. The density of the fluff must be also chosen with care; the higher the fluff density, the more material that will get in the way of any expanding star material. But if the fluff density is too low, the sound-crossing time of the fluff material will limit the timestep of the entire simulation. Ideally, the density of the fluff will be able to be orders of magnitude lower than that of the ejecta, but for a Type Iax this is difficult as parts of the sparse ejecta reach 10$^{-2}$ g/cm$^3$. Here, we set the fluff density to 10$^{-4}$ g/cm$^3$, and choose a refinement scheme based on the location of the fluff, rather than density.

\begin{figure}[ht]
\begin{center}
\includegraphics[scale=0.8]{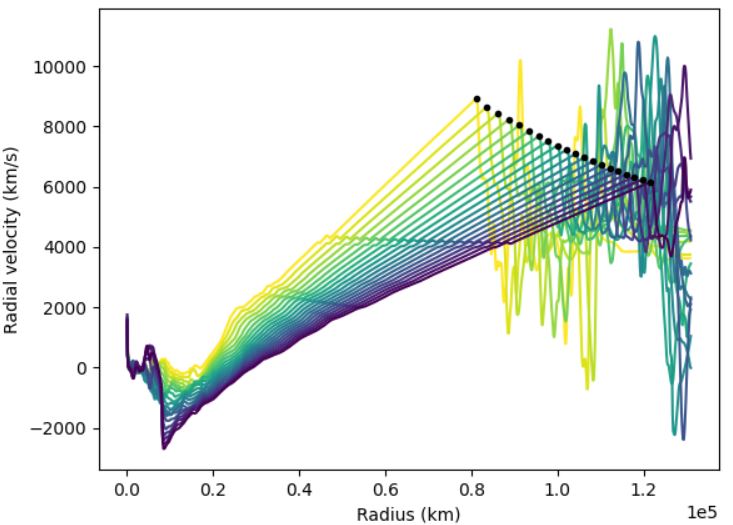}
\end{center}
\caption{\label{fig:vrad} Radial velocity vs.\ radius of our simulation over time. Each line represents an individual timestep, from 10.1 s (yellow) to 20 s (purple) in intervals of 0.5 s. Time follows the darkening colors. The black points mark the highest velocity that the ejecta reaches at each timestep, which decreases with time due to interaction with the fluff.}
\end{figure}

Figure \ref{fig:vrad} shows the radial velocity profile of one deflagration simulation over time. At small radii, the bound remnant pulls material back towards the center, giving it an increasingly smaller radial velocity. At larger radii, the material becomes unbound and will be ejected from the system. Notably, in the ejecta, the characteristic linear relationship between the radial velocity and the radius can be seen, demonstrating that the ejecta are in free expansion. Past the point of the ejecta, the ejecta and fluff interact, leading to a noisy radial velocity profile. 

At the ejecta-fluff boundary, the ejecta pushes forward and sends a shock into the fluff, which responds by piling up at the boundary and sending a reverse shock back into the ejecta, slowing it down. The ejecta has enough energy that it is able to overtake the fluff and continue pushing it aside, moving outwards. Therefore, unlike true free expansion where over time the velocity of the ejecta would remain unchanged, the ejecta loses speed. This is most apparent at the boundary between the ejecta and the fluff, denoted by the black points in Figure \ref{fig:vrad}, where the velocity of the ejecta at the boundary decreases with time. The slope of the line, which is the inverse of the expansion time, also decreases, signifying that the expansion time is increasing. Therefore, interaction with the fluff is a limiting factor in how long we can run our simulations until they are accurate.

One way to ameliorate the effects of the fluff is the implementation of a ``roll-off'', where the density and temperature of the fluff is linearly decreased with radius. This strategy is implemented and discussed in \cite{Boos_2021}. The higher fluff density at the star's surface will both prevent the timestep from being constrained and make it easier to keep the star-fluff boundary in pressure equilibrium without the need to drastically increase the fluff temperature. At the same time, the roll-off will provide overall less material for the expanding ejecta to run into, reducing the strength of the reverse shock and enabling the ejecta to be followed further in time. \cite{Kromer_2015} and other studies follow the ejecta out to 100 s or more, and a fluff roll-off could make this feasible for our and other future studies.

\subsection{Reignition} \label{sec:reignition}

\begin{figure}[hb]
\centering
\includegraphics[scale=0.17]{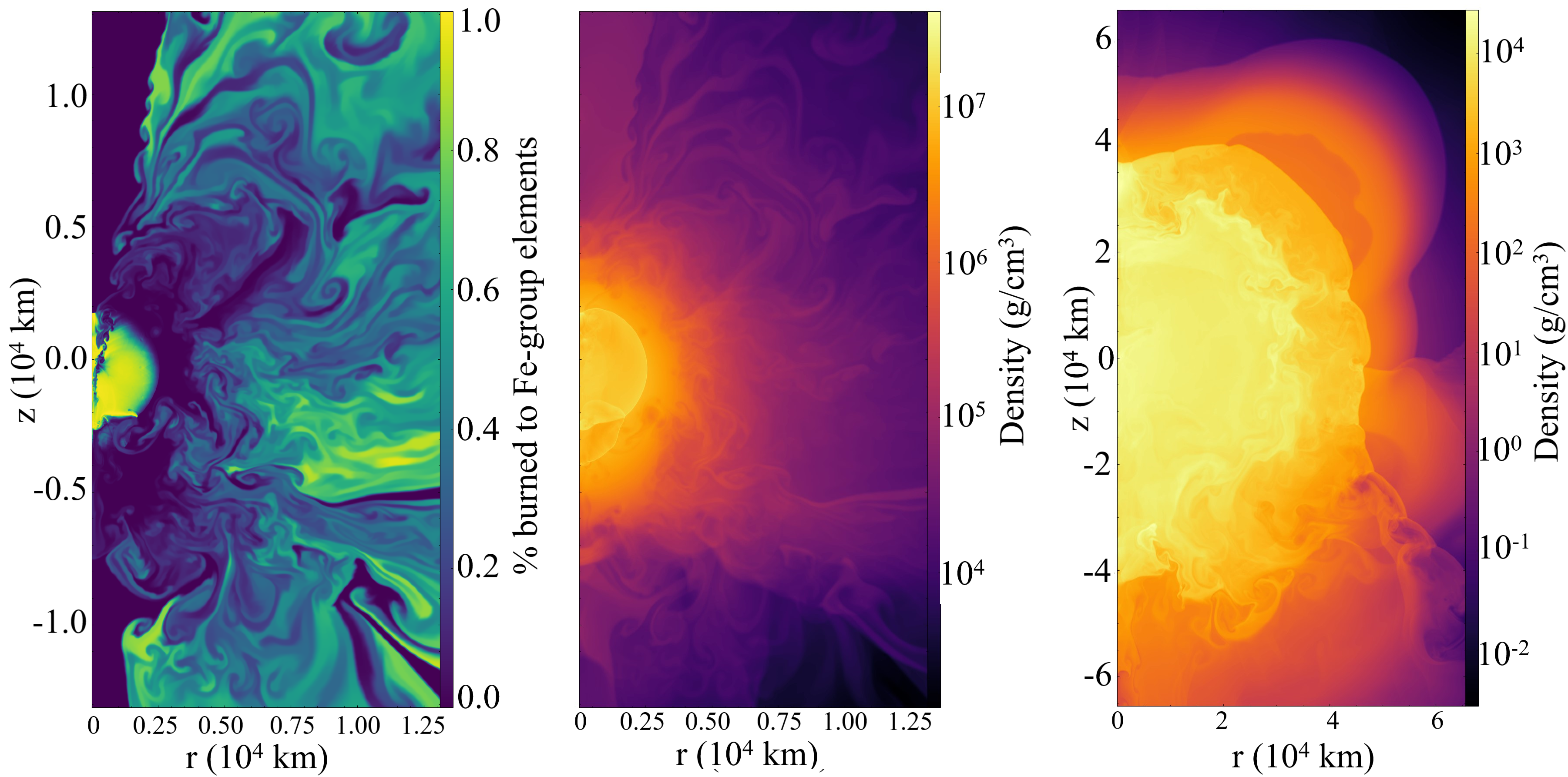}
\caption{\label{fig:reignition}Reignition of the remnant caused by material fallback  in realization 28. (Left) The percent of material burned to Fe-group elements at 7.3 s. The detonation can be seen in the center of the remnant, which is mostly unburned fuel surrounded by plumes of burned material from the deflagration phase. (Center) The density at 7.3 s. The detonation can be seen in the center of the remnant, which is much denser than the surrounding plumes. (Right) The density at 12 s, where the expanding burned material and the detonation shock can be seen.}
\end{figure}

As we ran out to the relatively long time of 20 s, some of our realizations reignited by means of a pulsational detonation caused by material falling back onto the dense, mostly unburned remnant. This process is visually demonstrated in Figure \ref{fig:reignition}. 
To prevent reignition from happening in our deflagration study, we manually turned off
burning at 5 s, by which time the deflagration had quenched. This manual setting is also used by other deflagration studies \cite{Lach_2021}, and a separate, accompanying study of pulsational detonations is sometimes performed \cite{Lach_2022_reignition}.

\begin{figure}[h]
\begin{minipage}{19pc}
\includegraphics[width=19pc]{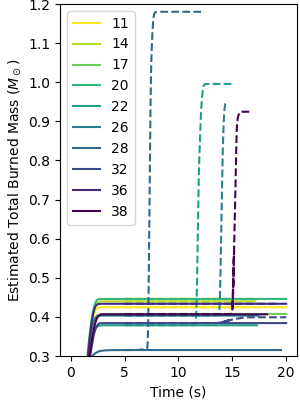}
\end{minipage}
\begin{minipage}{19pc}
\includegraphics[width=19pc]{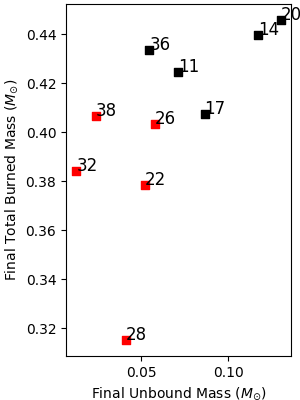}
\end{minipage} 
\caption{\label{fig:reignition_yields} (Left) Burned mass vs.\ time for selected realizations run to test reignition. The solid lines show the results from our deflagration study (burning manually turned off at 5 s), and the dotted lines show the results for thermally-activated burning turned on after the deflagration completes. (Right) Total burned mass vs.\ total ejected mass of selected realizations from our deflagration study. Marked in red are the realizations that reignited, and black are the ones that did not (yet).}
\end{figure}

This reignition is interesting and prompts many questions. Why do some, but not all, the realizations reignite before 20 s? Or do all realizations reignite, but at different times? Does reignition depend on the amount of mass ejected from the system? To probe these questions, we artificially reduced the flame speed to 0 after the deflagration completed so that the flame would essentially be deactivated; the ADR scheme is tuned to approximate the initial deflagration phase of an SN Ia, and not suited for modeling reignition. Instead, the thermally-activated C-burning reaction rate is used in place of the ADR equation to evolve the first (C-burning) stage of our burning model. The results from a few selected realizations that span the space of ejected mass are shown in Figure \ref{fig:reignition_yields}.

As expected, realizations that ejected less mass in our deflagration study reignite sooner, most likely because the remnant is more massive, denser, and hotter, which makes it easier for the fallback to create a spark. In general, they also burn more mass in the reignition, because there is more mass at higher densities available to consume. It cannot be definitively determined whether or not all realizations will reignite; the process may just take much longer for some than for others. 

It is also important to note that our simulation is unsuited to determine conditions for initiating a pulsational detonation, and therefore this process has a great deal of uncertainty. These results are interesting, but absolutely not methodical; they simply showcase an idea to be rigorously tested. First, it would need to be determined what resolution properly describes the uneven surface of the bound, compact remnant, the structure that detonates. Secondly, our three-stage burning model was created for a flame front moving through relatively homogeneous regions of fuel. In this case, the detonation occurs in a heterogeneous environment -- the C/O fuel of the initial WD is mixed with burning ashes -- and it would need to be determined if our burning model is still applicable. An additional, but certainly not the last, question is whether the C-burning reaction alone is enough to accurately describe the ignition, or if the burning is enhanced or suppressed by another reaction that would have to be included.

\subsection{Initial Flame} \label{sec:initflame}
As the sub-cm scales of a realistic flame cannot be resolved in our star-scale simulation, a deflagration is often ignited by inserting a volume of fully-burned material into the star. The size, location, and number of the initial flame bubbles are often varied to determine its effect on the simulation results. A smaller flame has longer to evolve and be affected by fluid instabilities before it reaches the size of a larger flame. However, it is unclear whether in reality ignition occurs in one location or multiple merging locations. Despite the uncertainties, it is important to vary this initial condition to determine its influence on the outcome of the simulation.

\begin{figure}[ht]
\begin{center}
\includegraphics[scale=0.15]{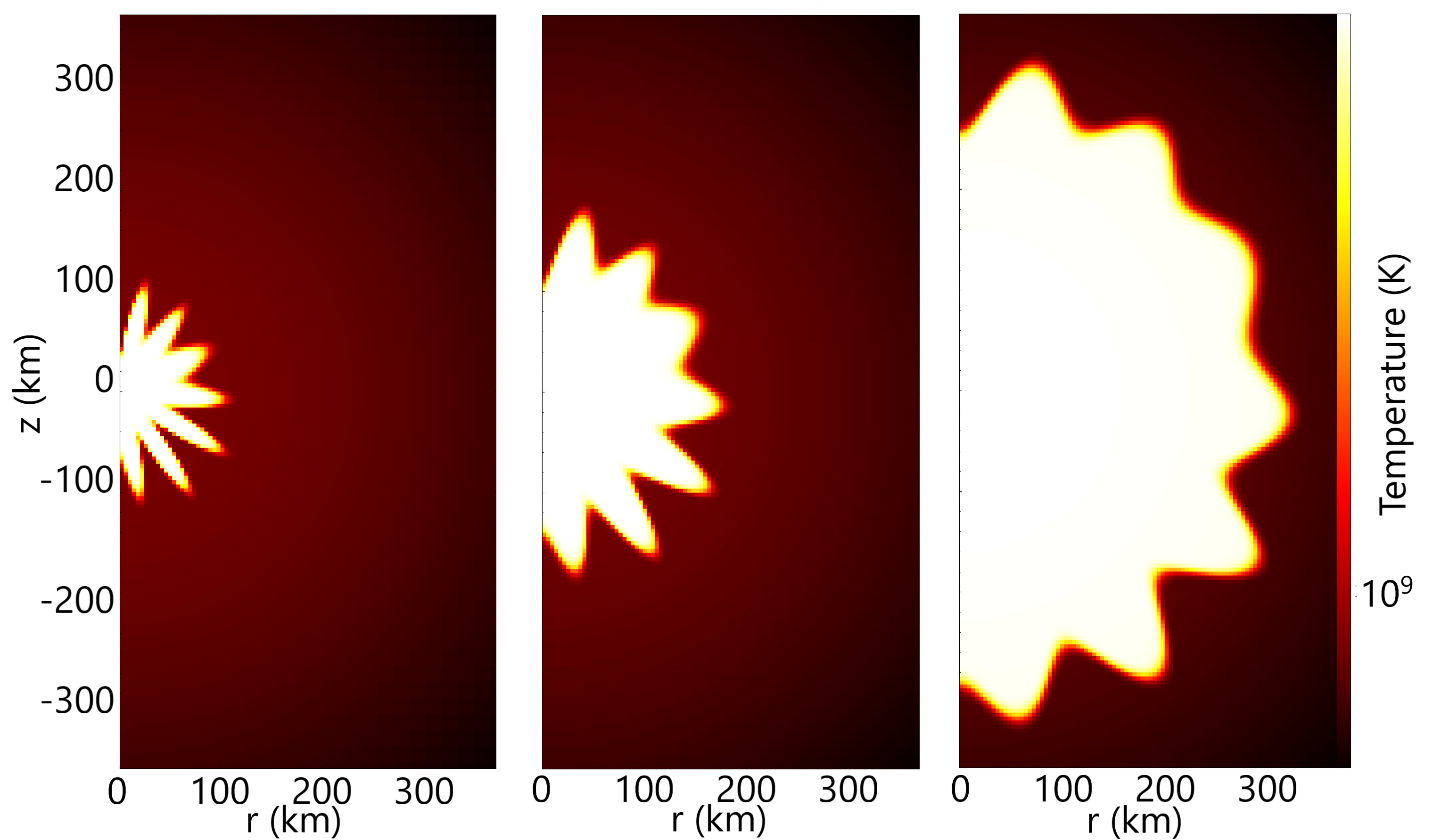}
\end{center}
\caption{\label{fig:initflame} Three initial flames from the same realization with (from left to right) 75 km, 150 km, and 300 km initial radii.}
\end{figure}

The initial flame in our simulation is placed at the temperature peak in the center of the star. It was given a radius of 150 km, a choice prompted by previous DDT studies \cite{Townsley_2007, Townsley_2009}. An 150 km initial flame allowed the star to expand enough during the deflagration phase so that the subsequent detonation produced yields consistent with observations. For our deflagration study, we investigated how the initial flame radius affected the results. We selected ten realizations that spanned the space of ejected $^{56}$Ni mass and adjusted their initial flames to 75 km and 300 km. Examples of these three flame radii for the same realization are shown in Figure \ref{fig:initflame}. Note that the radii and not the perturbations were scaled for this experiment.

\begin{figure}[ht]
\begin{center}
\includegraphics[scale=0.8]{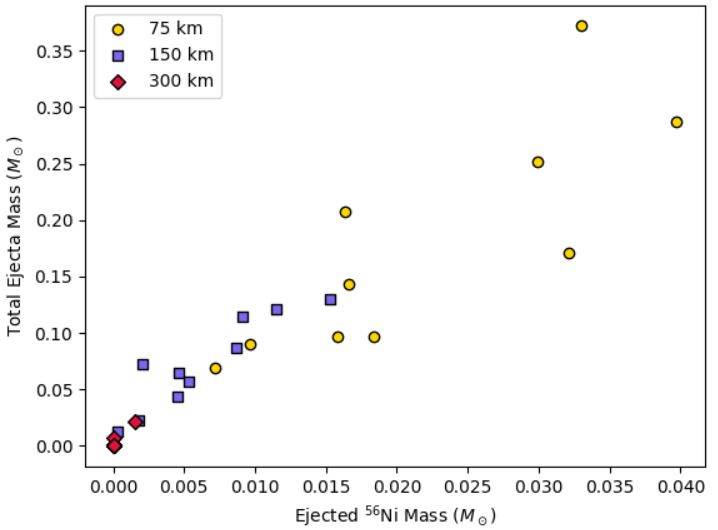}
\end{center}
\caption{\label{fig:initflame_yields} Total ejected mass vs.\ ejected $^{56}$Ni mass for ten realizations using three different initial flame radii: 75 km (yellow \opencircle), 150 km (purple \opensquare), and 300 km (pink \opendiamond). In general, there is a direct relationship between the amount of ejected mass and the amount of ejected $^{56}$Ni. Smaller initial flames produce more ejecta. Eight of the ten realizations with the 300 km radius initial flame produced no ejecta.}
\end{figure}

Figure \ref{fig:initflame_yields} shows the total ejected mass and the total ejected $^{56}$Ni mass for the three initial flame radii. There is a clear linear relationship between the ejected mass and the ejected $^{56}$Ni mass; roughly 10\% of the ejecta is $^{56}$Ni. In general, a larger initial flame radius produces less ejected mass, and less burned mass overall. This is because the larger flame injects more energy into the simulation at its start, causing the star to expand faster. Therefore, the burning occurs at lower densities and extinguishes faster, leading to a smaller burned yield and less overall energy release. Perhaps most interestingly, eight of the ten 300 km radius realizations produced no ejected mass -- all of the mass remained gravitationally bound. Based on the findings of Section \ref{sec:reignition}, the 300 km radius realizations are likely to all reignite, while the 75 km ones are more likely to not (at least within 15-20 s).

Importantly, decreasing the flame radius still produces ejected $^{56}$Ni masses within the range of observations of SNe Iax, meaning that the deflagration model remains a viable hypothesis. If reignition does not occur, a deflagration and resulting SN Iax could also be a potential explanation for hypervelocity WDs such as LP 40-365 \cite{ElBadry_2023}.

\vspace{-12pt}
\section{Summary and Conclusions} \label{sec:summary}
Numerical models provide a means to experiment on astrophysical systems. These tools comes with challenges including choosing initial conditions, assumptions, and approximations. We discuss the challenges faced when simulating the dim SNe Iax as a deflagration in a WD. We find that the slow-moving, sparse, and cool ejecta are sensitive to the choices made in simulation: the EOS ventures into an invalid region when describing it; the fluff impedes its progress; and the choice of the burning model and the initial flame size determines its mass. Despite all of these uncertainties, our simulation demonstrates that a deflagration in a WD is a viable mechanism to produce an SN Iax, encouraging future exploration in 3D.

It is important to understand the variability of simulations under different initial conditions and assumptions, especially for non-linear processes. Performing an extensive parameter study for a detailed simulation might be computationally prohibitive, however testing more abstracted simulations such as the 2D suite performed here can provide insight into a model's sensitivity.

\ack
The authors would like to thank Stony Brook Research Computing and Cyberinfrastructure, and the Institute for Advanced Computational Science at Stony Brook University for access to the SeaWulf computing system, which was made possible by a \$1.4M National Science Foundation (NSF) grant (\#1531492). This research was also supported in part by the US Department of Energy (DOE) under grant DE‐FG02‐87ER40317, in part by the NSF under grant OAC 1927880, and in part by the Simons Summer Research Program at Stony Brook University. \\

\noindent \textbf{References}
\bibliography{bibliography}

\end{document}